\documentclass[fleqn,10pt]{wlscirep}
\usepackage{graphicx}
\usepackage{amssymb}
\usepackage{epstopdf}
\usepackage{url}
\usepackage{subfigure}
\usepackage{lineno}
\usepackage{color}

\usepackage{mathtools}
\usepackage{color, colortbl}
\usepackage{fancyhdr}
\usepackage{ctable}
\definecolor{Gray}{gray}{0.9}
\usepackage{enumitem}
\usepackage{graphicx}
\usepackage{caption}
\usepackage{color, colortbl}
\usepackage{array}
\usepackage{booktabs} 
\usepackage{makecell}
\usepackage{cellspace}
\usepackage{tabularx}

\title{Multidimensional Tie Strength and Economic Development}

\author[1,2*]{Luca Maria Aiello}
\author[3]{Sagar Joglekar}
\author[3,4]{Daniele Quercia}
\affil[1]{IT University, Copenhagen, 2300, Denmark}
\affil[2]{Pioneer Centre for AI, Copenhagen, 2100, Denmark.}
\affil[3]{Nokia Bell Labs, Cambridge, CB30FA, United Kingdom}
\affil[4]{CUSP, King's College London, WC2R2LS, United Kingdom}
\affil[*]{luai@itu.dk}


\begin{abstract}
The strength of social relations has been shown to affect an individual's access to opportunities.  To date,  however, the correspondence between tie strength and population's economic prospects has not been quantified,  largely because of the inability to operationalise strength based on Granovetter's classic theory.  Our work departed from the premise that tie strength is a unidimensional construct (typically operationalized with frequency or volume of contact), and used instead a validated model of ten fundamental dimensions of social relationships grounded in the literature of social psychology.  We built state-of-the-art NLP tools to infer the presence of these dimensions from textual communication,  and analyzed a large conversation network of 630K geo-referenced Reddit users across the entire US connected by 12.8M social ties created over the span of 7 years.  We found that unidimensional tie strength is only weakly correlated with economic opportunities ($R^2=0.30$),  while multidimensional constructs are highly correlated  ($R^2=0.62$).  In particular,  economic opportunities are associated to the combination of: \emph{i)} knowledge ties,  which bridge geographically distant groups, facilitating the  knowledge dissemination across communities; and \emph{ii)} social support ties,  which knit geographically close communities together, and represent dependable sources of social and emotional support.  These results point to the importance of developing high-quality measures of tie strength in network theory. 
\end{abstract}

\begin{document}

\flushbottom
\maketitle
\thispagestyle{empty}

\section*{Introduction}

The strength of social relations has been shown to affect an invididual's access to innovation~\cite{rogers2010diffusion},  access to economic opportunities~\cite{granovetter2005impact}, life expectancy~\cite{holt2010social}, and happiness~\cite{fowler2008dynamic}. According to Granovetter’s classic theory about tie strength~\cite{granovetter1977strength}, information flows through social ties of two strengths. First, through weak ties. These ties, despite being used infrequently, bridge distant groups that tend to posses diverse information, facilitating the knowledge dissemination across communities. Second, information also flows through strong ties. These ties, by being used frequently,  knit close communities together,  and represent dependable sources of social and emotional support. 

To date, however, the correspondence between tie strength and population's economic prospects has not been quantified,  largely because of the inability to operationalize tie strength based on Granovetter's conception. Typically, network studies operationalize strength with indicators like frequency or volume of contact~\cite{marsden12}. Eagle \emph{et al.} did so by studying the relationship between the structure of a national communication network and access to socio-economic opportunity~\cite{eagle2010network}. They found that network diversity was associated to opportunities, but communication volume or number of contacts was not.  The prospect that tie strength is not a unidimensional construct ranging from weak to strong but might be multidimensional is broadly consistent with theoretical and experimental work by Marsden and Campbell~\cite{marsden12} and Wellmann and Wortley~\cite{wellman90different}.   It is also consistent with Granovetter's orginal operationalization of strength as \emph{``a (probably linear) combination of the amount of time,  the emotional intensity,  the intimacy (mutual confiding),  and the reciprocal services which characterize the tie.''}~\cite{granovetter1977strength} These indicators have been repeatedly found to be only weakly related to frequency of contacts~\cite{marsden12,eagle2010network}. Therefore, network studies using frequency of contacts to model strength are capturing only one aspect of the linkages among individuals. 

Our work departed from the premise that tie strength is a unidimensional construct, built upon work on social psychology starting from Granovetter's conception of tie strength, and identified and validated ten fundamental \emph{dimensions} of social relationships~\cite{deri18coloring,choi20social}. In previous work, we showed that these ten dimensions correspond to how people perceive and categorize most of their own social relationships~\cite{deri18coloring}, and we built a state-of-the-art NLP tools to infer the presence of these dimensions from textual communication~\cite{choi20social}. In this work, we used these tools to analyze a large conversation network of geo-referenced Reddit users across the entire US ($\sim$13M ties). Then, going back to Eagle \emph{et al.}'s work and borrowing their methodological framework~\cite{eagle2010network}, we were able to test whether the structure of a national communication network (in particular, its tie diversity) was related to access to socio-economic opportunities,  and whether switching from a unidimensional notion of tie strength to a multidimensional one would improve explanatory power. We found that tie diversity measured on the networks of knowledge exchange and social support correlates much more strongly with economic development ($R^2=0.62$) than diversity measured on a network simply weighted on frequency of interactions ($R^2=0.30$). 

In line with Granovetter's conception of tie strength, we found that knowledge ties and social support ties: are hardly distinguishable solely based on frequency of interaction;  have opposite geographic distribution (knowledge ties are global,  spanning longer geographical distances, while social support ones are local,  typically staying in the same state); and both contribute to economic opportunities (states with higher GDP per capita are characterized by both global access to knowledge and local access to support). These results point to the importance of developing multidimensional measures of tie strength in network theory to better reflect the nature of human relationships that social links ought to model.

\section*{Results}

From a set of 65M comments posted on Reddit by 1.3M users between the years of 2006 and 2017, we extracted the social interactions of all Reddit users that we could geo-reference at the level of the 51 US states using high-accuracy heuristics validated in previous work (see Methods). In Reddit, conversations develop over discussion \emph{threads}. If user $i$ commented over either a submission or a comment of another user $j$, we considered that $i$ sent a \emph{message} to $j$, as it is common practice when studying Reddit conversation networks~\cite{choi2015characterizing}. We created a directed communication graph $\mathcal{G}(U,E)$ to model such exchange of messages. The set of nodes $U$ contains all the geo-referenced Reddit users in our dataset. Two users $i$ and $j$ are connected by a directed edge $(i, j, w(i, j)) \in E$ if user $i$ sent at least one message to user $j$. The edge weight $w(i, j)$ represents the frequency of contacts and it is equal to the total number of messages sent. In total, the graph contains 630K nodes and 12.8M edges. The distribution of node degree and link strength is shown in Figure~SI1.

By applying our social dimensions classifier to the corpus of messages, we identified the subset of messages that express a social dimension $d$ (see Methods for details). In particular, we focused on the dimensions of \emph{knowledge exchange} and \emph{social support} (respectively, \emph{knowledge} and \emph{support} for short). Other dimensions are discussed in Supplementary Information). The classifier ranked the messages according to their likelihood of containing expressions of a given social dimension; we marked with dimension $d$ only the top 1\% of messages from the likelihood ranking of $d$ (we discuss results with looser thresholds in Supplementary Information, Figure~SI2). Out of these smaller sets of messages, we constructed \emph{dimension-specific communication graphs} $\mathcal{G}_d$ using the same procedure we adopted for building the overall communication graph $\mathcal{G}$. Such dimension-specific graphs capture only one type of social interaction each; for example, the \emph{knowledge} graph $\mathcal{G}_{knowledge}$ contains only edges formed by knowledge-exchange messages, and edge weights encode the number of knowledge-exchange messages flowing between the two endpoints. The dimension-specific graphs contain roughly $1\%$ of the edges of the full communication graph and between 16\% to 23\% of its nodes, depending on the dimension (see Table~\ref{table:literature_review}). The networks of \emph{knowledge} and \emph{support} include 20\% and 21\% of all nodes, respectively. The edges of $\mathcal{G}_{knowledge}$ and $\mathcal{G}_{support}$ overlap only slightly: around 2\% of the edges of each graph are also present in the other.

\begin{figure*}[t!]
    \centering
    \includegraphics[width=0.99\linewidth]{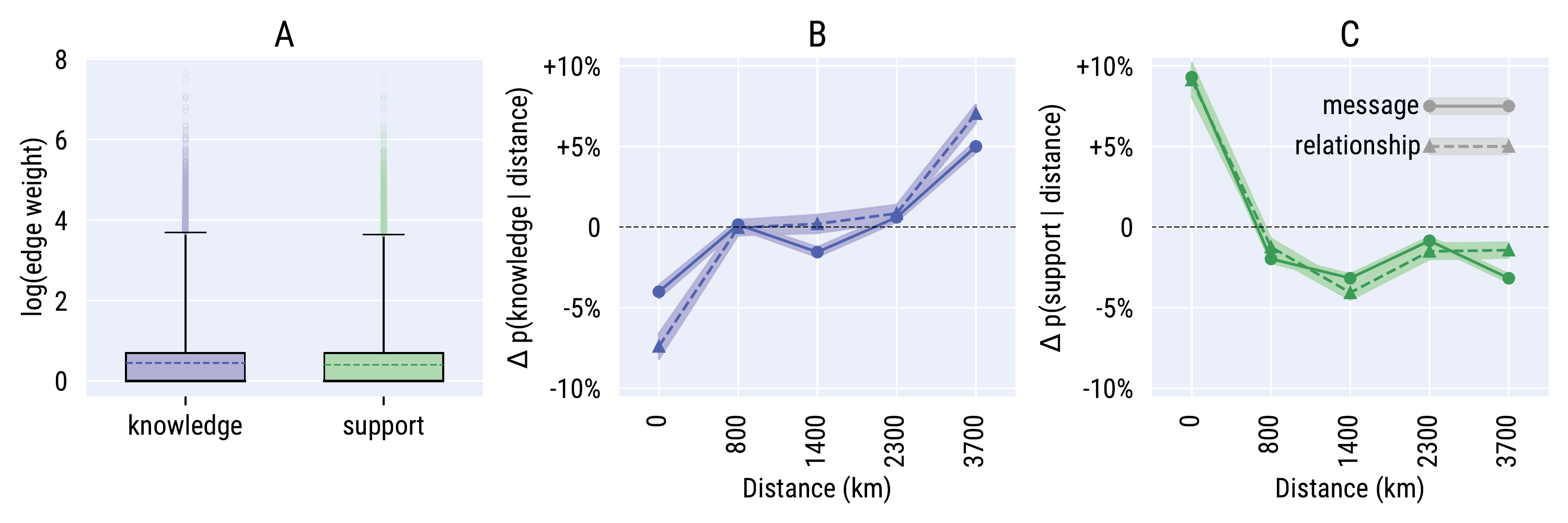}
		\caption{A) Boxplots of the weight distributions of ties that exchanged at least one message of \emph{knowledge} or one message of \emph{support}, on a logarithmic scale. Boxes represent the two mid quartiles of the distributions, with the median marked with a dashed line. The whiskers show the 99th percentiles of the distributions. B, C) Percent change $\Delta p(d|l)$ of the probability that a dimension $d$ is expressed by a social tie spanning a geographical distance $l$, compared to random chance. The change is estimated by comparing the real data with distance measurements on 50 instances of a null model that reshuffled user locations at random; the average values along with their $95\%$ confidence intervals are reported. Distances are discretized in five bins, each containing the same number of social ties. Bins are labeled with the median distance of the ties inside that bin. The `zero distance' bin contains almost exclusively pairs of users who live in the same state. Two types of measurements are presented: i) at the level of social relationships, where each social tie is counted once regardless of its weight, and ii) by performing a distance measurement for each individual message, thus effectively weighting more pairs of users who communicated frequently.}
    \label{fig:dim_vs_dist}
\end{figure*}

By having a sample of edges annotated with both social dimensions and weight, we were able to look into the relationship between frequency of contacts, \emph{knowledge}, and \emph{support}. The typical weight of edges connecting users who exchange knowledge is not dissimilar from the typical weight of those providing support. Figure~\ref{fig:dim_vs_dist}A compares the weight distribution of edges connecting users who exchanged knowledge with the weight distribution of edges connecting those who exchanged support. A two-sample Kolmogorov-Smirnov test (a statistic to measure the distance between two distributions) indicated that the two distributions, albeit statistically different, are very similar: $KS = 0.03$~$(p=0.0)$ on a range from 0 (indicating identical distributions) to 1 (maximum difference). This comparison exposes the inherent limit of quantifying tie strength with the mere frequency of interactions to adequately qualify the nature of social relationships.

In Reddit conversations, the main difference between \emph{knowledge} and \emph{support} ties does not lie in their strength but in their geographic span. The probability of creating \emph{knowledge} ties increases with the geographical distance between the two endpoints, while the probability of creating \emph{support} ties drops with distance (Figure~\ref{fig:dim_vs_dist}B,C). This is consistent with theoretical expectations. Knowledge production on the Web follows Pareto's law: a restricted number of experts create and spread information to a vast audience~\cite{baeza2015wisdom}; consequently, knowledge ends up being locally scarce~\cite{aral2011diversity} and needs to travel longer distances to reach multiple communities. In past studies, a similar pattern was detected for the communications within large corporations, where geographically distant ties were estimated to be more effective conduits for knowledge flow~\cite{reagans2003network,bell2007geography}. The opposite trend holds for \emph{support}. Geographical distance impacts significantly people's ability to provide both material and emotional support~\cite{mok2007did}. Despite computer-mediated communication has grown the opportunities for providing remote support~\cite{hampton2001long}, people have an innate sense for local attachments and an economic advantage to foster them~\cite{mesch1998social}, which might be why \emph{support} appears more rarely in long-distance relationships~\cite{wellman90different}.

Last, we tested if dimension-specific graphs are more indicative of economic development than the full communication graph. We did so by borrowing the experimental setup by Eagle \emph{et al.}~\cite{eagle2010network}, who studied the network of phone calls among residents of England and measured the \emph{spatial and social diversity} ($D_{spatial}$) for each of nearly 2,000 regional exchanges in the country. $D_{spatial}$ captures the diversity of areas that the residents of a given area communicate with, and they found it to be correlated with the Index of Multiple Deprivation---a composite score of social and economic development based on UK census data. They also tested the robustness of their results with an alternative measure of diversity $D_{social}$ that captures the diversity of people connected to the residents of a given area. We reproduced Eagle et al.'s experimental setup and ran an Ordinary Least Squares linear regression (OLS) to predict per-capita Gross Domestic Product (GDP) of US states in the year 2017~\cite{gdp_data} from the spatial diversity at state-level computed on \emph{i)} the full communication graph ($D_{spatial}$) and \emph{ii)} the two dimension-specific communication graphs ($D^{knowledge}_{spatial}$, $D^{support}_{spatial}$). Results for $D_{social}$ are highly aligned with those for $D_{spatial}$, and we discuss them in Supplementary Information. We focused on 44 states for which Reddit penetration is sufficient and aligned with the population distribution (see Methods), however we found qualitatively similar results when considering all states (see Supplementary Information, Table~SI3). Regressions models with different combinations of social and spatial diversity are presented in Tables~SI1~and~SI2.

In Table~\ref{tab:regression} we compare three linear regressions models: one based on population density only (a validated predictor of economic growth~\cite{bettencourt2013origins}), one using spatial diversity on the full graph with links weighted based on frequency of interaction, and one using the two spatial diversity scores calculated on the graphs of \emph{knowledge} and \emph{support}. The model based on the selected social dimensions is 138\% more accurate than the density-only baseline, while the model based on the full communication graph is only 15\% more accurate. To check whether the difference in performance is due to the selection of \emph{knowledge} and \emph{support} ties or just to the smaller sample considered, we ran a regression using a random sample of ties as small as the number of \emph{knowledge} ties, and obtained the worst fit ($R^2_{adj}$ of approximately $0.1$, see Supplementary Information).

In the regression model with the social dimensions, the coefficient for \emph{knowledge} diversity is positive and the one for \emph{support} diversity is negative. People living in areas characterized by superior economic outcomes access novel information that is not available locally by establishing a diverse set of global interactions, which is in agreement with the \emph{weak tie} pillar of Granovetter's theory. Residents of states with highest per-capita GDP draw their social support mostly from local connections, in agreement with the \emph{strong tie} pillar of the theory. The effect size of \emph{knowledge} is stronger (almost double) than the effect size of \emph{support}, which indicates that the process of knowledge exchange is the primary correlate of economic development, and the network of support compounds over it. A linear regression including other social dimensions is discussed in Table~SI4, but the interplay between \emph{knowledge} and \emph{support} is more predictive than any other combination of dimensions.

{\def\arraystretch{1.5}
\begin{table}[!t]
\footnotesize
\setlength{\tabcolsep}{5pt}
\begin{center}
\begin{tabular}{lccc}
\multicolumn{4}{c}{\textbf{}}\\
\specialrule{.1em}{.05em}{.05em} 
\multicolumn{4}{c}{\textbf{Population density}}\\
\textbf{Feature}	& \textbf{$\beta$} &	\textbf{SE} & \textbf{$p$} \\
\hline
$\alpha$ (intercept)	           & 0.310  & 0.045  & 0.000 \\
Pop. density                     & 0.636  & 0.113  & 0.000 \\
&  &  &  \\
&  &  &  \\
\hline
\multicolumn{2}{l}{Durbin-Watson stat. = 1.982} 	& \multicolumn{2}{l}{\textbf{$R_{adj}^2$ = 0.26}}   \\
\specialrule{.1em}{.05em}{.05em} 
\end{tabular}
\quad
\begin{tabular}{lccc}
\multicolumn{4}{c}{\textbf{Predicting GDP per capita from:}}\\
\specialrule{.1em}{.05em}{.05em} 
\multicolumn{4}{c}{\textbf{Diversity on full communication graph}}\\
\textbf{Feature}	& \textbf{$\beta$} &	\textbf{SE} & \textbf{$p$} \\
\hline
$\alpha$ (intercept)	           & -0.035  & 0.108  & 0.747 \\
Pop. density                     & 0.565  & 0.174  & 0.002 \\
$D_{spatial}$                    & 0.243 & 0.151  & 0.116 \\
                   &  &   & \\
\hline
\multicolumn{2}{l}{Durbin-Watson stat. = 2.082} 	& \multicolumn{2}{l}{\textbf{$R_{adj}^2$ = 0.30}}   \\
\specialrule{.1em}{.05em}{.05em} 
\end{tabular}
\quad
\begin{tabular}{lccc}
\multicolumn{4}{c}{\textbf{ }}\\
\specialrule{.1em}{.05em}{.05em} 
\multicolumn{4}{c}{\textbf{Spatial diversity on dimension-specific graphs}}\\
\textbf{Feature}	& \textbf{$\beta$} &	\textbf{SE} & \textbf{$p$} \\
\hline
$\alpha$ (intercept)	           & 0.1943  & 0.061  & 0.003 \\
Pop. density                     & 0.4713  & 0.116  & 0.000 \\
$D_{spatial}^{knowledge}$         & 1.0327  & 0.164  & 0.000 \\
$D_{spatial}^{support}$          & -0.5549 & 0.154  & 0.001 \\
\hline
\multicolumn{2}{l}{Durbin-Watson stat. = 2.069} 	& \multicolumn{2}{l}{\textbf{$R_{adj}^2$ = 0.62}}   \\
\specialrule{.1em}{.05em}{.05em} 
\end{tabular}

\end{center}
\caption{Linear regressions to predict GDP per capita of US states from: (left) population density only; (center) spatial diversity computed on the full communication graph; (right) spatial diversity computed on dimension-specific communication graphs. Population density is added as a control variable in the latter two models. Adjusted $R^2$ and Durbin-Watson statistic for autocorrelation (values close to 2 indicate no autocorrelation) are reported. The contribution of individual features to the models is described by their $beta$-coefficients, standard errors (SE) and $p$-values.}
\label{tab:regression}
\end{table}
}

\section*{Discussion} 

In agreement with Granovetter's theory, we found that economic development at the level of US states is associated to the abundance of global ties that carry factual knowledge and with the abundance of local ties providing social support. This finding is compatible with the established notion of innovation being fueled primarily by novel information flowing from diverse regions of the social network, and secondarily by an adequate support network to favor the re-elaboration of those ideas locally. This perspective enriches the corpus of experimental evidence about the existence of a trade-off between seeking novel information and building tight networks of support~\cite{aral2011diversity,aral2016future,rajkumar2022causal}. We showed that geographical regions generally experience that trade-off but the regions that achieve high economic success are those that have both global outreach of knowledge exchange and local networks of support.

In contrast with a variety of network science studies, we provided evidence that frequency of contacts might not be a good proxy for tie strength: network diversity calculated on a weighted social network is weakly associated to economic development at state level. Moreover, our results challenge the equivalence between weak ties and knowledge flow, at least for the case of Reddit. Interestingly, we found that \emph{knowledge} and \emph{support} ties differ in terms of their geographical span, with \emph{knowledge} ties being far-reaching, and \emph{support} ties being local. 

The ability of measuring directly these two aspects of social interaction that are postulated by Granovetter's theory to be drivers to innovation enhances the predictive and descriptive power of network models. Strikingly, narrowing down the analysis to a small subset of messages that express either \emph{knowledge} or \emph{support} yields a predictive performance that is as much as double of that of models used in previous research that considered only frequency of contacts~\cite{eagle2010network}.

The ability of decomposing relationship data into interpretable social constituents opens up ample avenues of exploration in social network analysis. Studying how different social dimensions are instantiated by different anatomic patterns of social networks such as their community structure or the centrality of their actors might be a promising research direction. Also, this work showed the association of \emph{knowledge} and \emph{support} with GDP, but other social dimensions may well explain other socio-economic outcomes such as health or quality of life.

Both our data and methods  suffer from limitations that future work may address. Unlike the work by Eagle et al., upon which our experimental setup was based~\cite{eagle2010network},  our study relies on social network data that covers only a small sample of the population; this was a necessary sacrifice in order to gain the crucial ability to analyze the content of social interactions.

Among all the social platforms from which we could have collected conversational text, we selected Reddit because its richness of information and variety of social interaction types. Other popular platforms (e.g., Facebook, Twitter) either authorize data collection exclusively from volunteer users~\cite{lambiotte2014tracking} or expose data APIs that may be limited by volume, temporal scope, and known sampling biases~\cite{morstatter2013sample}. On the contrary, Reddit allows for the collection of the full conversation history between any pair of users, and includes metadata useful for their characterization, such as geo-localization~\cite{medvedev2017anatomy}. Also, Reddit's etiquette, credit system, and topic-oriented subreddits encourage social participation for purposes that are akin to real-life social networks~\cite{anderson2015ask}, such as socialization, entertainment, and information exchange~\cite{moore2017redditors}, while naturally disincentivizing practices that disproportionately favor status-seeking, which are prominent in platforms such as Twitter and Facebook~\cite{park2009being,phua2017uses}. As a result, Reddit's comment threads enjoy properties that are typical of human conversations, such as the high topical coherence of successive messages in a thread~\cite{weninger2013exploration,choi2015characterizing}. Because of these desirable properties, Reddit has been the platform of choice for hundreds of quantitative and qualitative studies on social behavior in the last ten years~\cite{proferes2021studying}. Furthermore, the anatomy and dynamics of the Reddit conversation network exhibit properties that are in line with those of most social networks~\cite{newman2003social, leskovec2008microscopic, ugander2011anatomy}, which speaks to the potential of our findings to generalize to other contexts. These properties include broad distributions of the node degree and of the frequency of most user activities~\cite{weninger2014exploration,cauteruccio2020investigating,baowaly2022co} (see also Figure~SI1), marked community structure~\cite{soliman2019characterization}, assortativity~\cite{cauteruccio2020investigating}, and burstiness of interactions~\cite{wang2012user}. Nevertheless, Reddit user base is biased towards males (64\%) and young adults (36\% in the age range 18-29, 22\% in the range 30-49), and our study focuses entirely on US residents~\cite{statista22distribution}; therefore, replicating our analysis to multiple conversation networks is in order to corroborate the robustness of our results.

Within Reddit, our perspective on the ecosystem of social interactions is restricted by our focus on the physical space. In particular, the communication graphs include only a sample of all the existing edges,  namely those that connect users whose geo-locations could be estimated. This entails three main biases. First, the majority of interactions are left out of the picture, thus potentially reducing the predictive and descriptive power of our models. Second, the social links we considered were not randomly sampled, as they connect users who self-selected themselves to join geo-salient subreddits. Last, the limited resolution of the user spatial location (state-level) affected our ability to perform a finer-grained geographic analysis (e.g., at city level). To address these biases,  future work ought to consider social systems where a larger portion of users can be geo-referenced at a finer geographic resolution.

Even if our social dimensions classifiers were trained on Reddit data and were shown to achieve high accuracy (see Methods), their output is not error-free. To improve both precision and recall, a systematic error analysis and a fine-tuning of the model with additional training data would be in order. The ten social dimensions,  albeit more comprehensive than any existing model, do not exhaustively map all the possible elements that define social interactions. The concepts that these social dimensions encode are rather broad and encompass a rich spectrum of nuances.  The main goal of this work was to go beyond simple frequency of contacts as a proxy for tie strength, offering well-founded interaction archetypes  that could be explored and refined in the future. 

\section*{Methods} 

\subsection*{Reddit data collection}\label{sec:dataset:reddit}

Reddit is a public discussion website particularly popular in the United States where half of its user traffic is generated. Reddit is structured in an ever-growing set of independent \emph{subreddits} (1.2M at the time of writing) dedicated to a broad range of topics~\cite{medvedev2017anatomy}. Users can post new \emph{submissions} to any subreddit, and other users can add \emph{comments} to submissions or to existing comments, thus creating nested conversation \emph{threads}.

The vast majority of Reddit submissions and comments since 2007 is publicly available through the \texttt{pushshift.io} API~\cite{baumgartner2020pushshift}. For the purpose of this study, we gathered the content created in two temporal windows: from 2007 until the end of 2012, and for the whole year of 2017. The findings presented in the Result section were obtained using the data from these two windows jointly, but having at hand two collections from distinct time periods allowed us to study how data recency affects the ability to predict the desired outcome (see Supplementary Information, Figure~SI3). In total, we collected 65M comments from 1.3M users.

We restricted our study to users whom we could geo-reference at the level of US States. Although Reddit does not provide explicit information about user location, we used a location-estimation heuristic proven to be effective in previous work~\cite{balsamo2019firsthand}. We first identified 2,844 geo-salient subreddits related to cities or states in the United (\url{https://www.reddit.com/r/LocationReddits/wiki/faq/northamerica}). We assigned a user to a state if \emph{i)} they posted at least $n$ submissions or comments in subreddits related to that state, and \emph{ii)} 95\% or more of their comments and submissions posted to geo-salient subreddits were done in subreddits related to that state. The findings presented earlier were obtained with $n=3$; in Supplementary Information (Figure~SI4) we discuss results obtained by varying this threshold. Overall, we found 632k users who are likely to be located in one of the 51 US states. The number of users per state ranges from less than 1k (Wyoming) to 61k (California). In total, these users posted 16.2M comments in total (9.8M in 2007-2012, and 6.4 in 2017).

\subsection*{Filtering states by Reddit penetration}

\begin{figure}[ht!]
    \centering
		\includegraphics[width=0.60\linewidth]{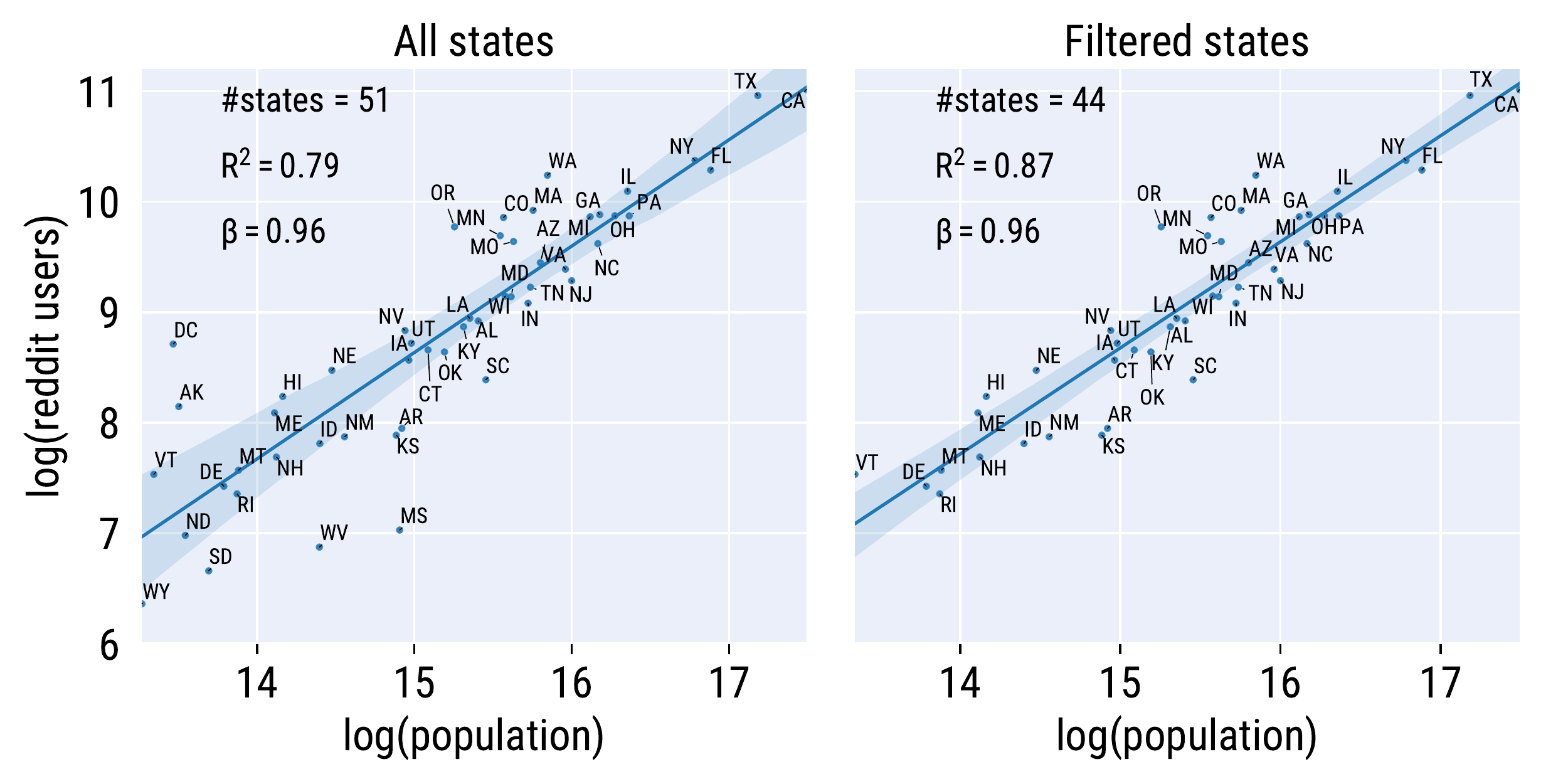}
		\caption{Relationship between population and number of Reddit users across US states. The best linear fit is shown, together with its slope $\beta$ and the $R^2$ coefficient to measure the goodness of fit. On the left, all states are included. On the right, the states whose Reddit penetration was too low or was not proportional to the population of residents were removed.}
    \label{fig:corr_population_userbase}
\end{figure}

States in which the number of Reddit users is not proportional to the number of residents might distort the representation of social communication patterns that actually take place in those states. To identify such cases, we proceeded as follows. We first plotted the census population in 2017 against the number of Reddit users, across states (Figure~\ref{fig:corr_population_userbase}, left). We then obtained the best linear fit of the data and calculated the residuals between the number of Reddit users and the predicted value according to the linear fit. Last, we calculated the distribution of residuals and removed states whose residuals were more than 1 standard deviation away from the average of the distribution. Those included two states whose Reddit user base was higher than what one would expect based on their population (DC and AK) and two for which it was lower (MS and WV). In addition, we removed three outlier states whose Reddit penetration was lowest (less than 1000 users), which left us with a total of 44 states (Figure~\ref{fig:corr_population_userbase}, right).

\subsection*{Social dimensions from textual conversations}\label{sec:dataset:dimensions}

{\def\arraystretch{1.5}
\begin{table}[ht!]
\centering
\footnotesize
\begin{tabular}{p{12mm} p{90mm} c}
\specialrule{.1em}{.05em}{.05em} 
\textbf{Dimension} & \multicolumn{1}{c}{\textbf{Description}} & \textbf{\% Nodes in $\mathcal{G}_d$} \\
\Xhline{2\arrayrulewidth}
Knowledge & Exchange of ideas or information; learning, teaching~\cite{fiske2007universal} & 0.20 \\
Support & Giving emotional or practical aid and companionship~\cite{fiske2007universal} & 0.21   \\
\hline
Power & Having power over the behavior and outcomes of another~\cite{blau64exchange} & 0.17 \\
Status & Conferring status, appreciation, gratitude, or admiration upon another~\cite{blau64exchange} & 0.22 \\
Trust & Will of relying on the actions or judgments of another~\cite{luhmann1982trust} & 0.23 \\
Romance & Intimacy among people with a sentimental or sexual relationship~\cite{buss2003evolution} &  0.22\\
Similarity & Shared interests, motivations or outlooks~\cite{mcpherson2001birds} & 0.21 \\
Identity & Shared sense of belonging to the same community or group~\cite{tajfel2010social} & 0.17\\
Fun & Experiencing leisure, laughter, and joy~\cite{argyle2013psychology} & 0.21 \\
Conflict & Contrast or diverging views~\cite{tajfel1979integrative} & 0.16 \\
\specialrule{.1em}{.05em}{.05em} 
\end{tabular}
\caption{The social dimensions of relationships surveyed by Deri at al.~\cite{deri18coloring}. The last column reports the fraction of nodes of the full communication graph $\mathcal{G}$ that are included in each dimension-specific graph $\mathcal{G}_d$. The fraction of nodes in the last column is not exclusive, because nodes can be found in multiple dimension-specific graphs. Our work focused mainly on the dimensions of \emph{knowledge} and \emph{support}.}
\label{table:literature_review}
\end{table}
}

Social science research proposed several categorizations of constitutional sociological dimensions that describe human relationships~\cite{fiske1992four,wellman90different,spencer2006rethinking}. By surveying such extensive literature, Deri et al.~\cite{deri18coloring}  compiled one of the most comprehensive categorizations to date, which identifies ten main \emph{dimensions} of social relationships (Table~\ref{table:literature_review}). This theoretical model is rather exhaustive in that most relationships are accurately defined by appropriate combinations of the ten dimensions---Deri et al. showed it by asking hundreds of volunteers to write down keywords that described their relationships and found that all of them fitted into the ten dimensions. The ten social dimensions are frequently expressed through conversational language and, most importantly, these verbal expressions can be captured with computational tools.

We infer the social dimensions from Reddit messages using the NLP model proposed by Choi et al.~\cite{choi20social}, which comes with a publicly-available python implementation (\url{http://www.github.com/lajello/tendimensions}). Given a textual message $m$ and a social dimension $d$, the model estimates the likelihood that $m$ conveys $d$ by giving in output a score from 0 (least likely) to 1 (most likely). Rather than using a multiclass classifier, the model includes ten independently-trained binary classifiers $C_d$, one per each dimension. This choice was driven by the theoretical interpretation of the social dimensions~\cite{deri18coloring}, as any sentence may potentially convey several dimensions at once (e.g., a message expressing both trust and emotional support). Each classifier is implemented using a Long Short-Term Memory neural network (LSTM)~\cite{hochreiter1997long}, a type of Recurrent Neural Network (RNN) that is particularly effective in modeling both long and short-range semantic dependencies between words in a text, and it is therefore widely used in a variety of NLP tasks~\cite{sundermeyer2012lstm}. Like most RNNs, LSTM accepts fixed-size inputs. This particular model takes in input a 300-dimension embedding vector of a word, one word at a time for all the words in the input text. Embedding vectors are dense numerical representations of the position of a word in a multidimensional semantic space. Such representations are learned from large text corpora. This model uses GloVe embeddings~\cite{pennington2014glove} learned from Common Crawl, a text corpus containing 840B tokens.

The dimensions classifiers $C_d$ were trained using about 9k sentences that were manually labeled by trained crowdsourcing workers. Most of these sentences were taken from Reddit, which makes it the ideal platform to apply the model on. In their experiments, Choi et al. reported very high classification performance which averages to an Area Under the Curve (AUC) of $0.84$ across dimensions, and specifically $0.82$ for \emph{knowledge} and $0.83$ for \emph{support}. AUC is a standard performance metric that assesses the ability of a classifier to rank positive and negative instances by their likelihood score, independent of any fixed decision threshold. The AUC of a random classifier is expected to be 0.5, whereas the maximum value is 1.

Given in input a message $m$, the classifier outputs a score $s_d(m)$ that expresses the likelihood that message $m$ contains dimension $d$. In practice, the classifier estimates a score for each sentence in $m$ and returns the maximum score, namely: $s_d(m) = \max_{sentence \in m} s_d(sentence)$. By using the maximum score, we considered a message as likely to express dimension $d$ as its most likely sentence, thus avoiding the dilution effect of the average. This reflects the theoretical interpretation of the use of the social dimensions in language~\cite{deri18coloring}: a dimension is conveyed effectively through language even when expressed only briefly.

\begin{figure}[t!]
    \centering
		\includegraphics[width=0.60\linewidth]{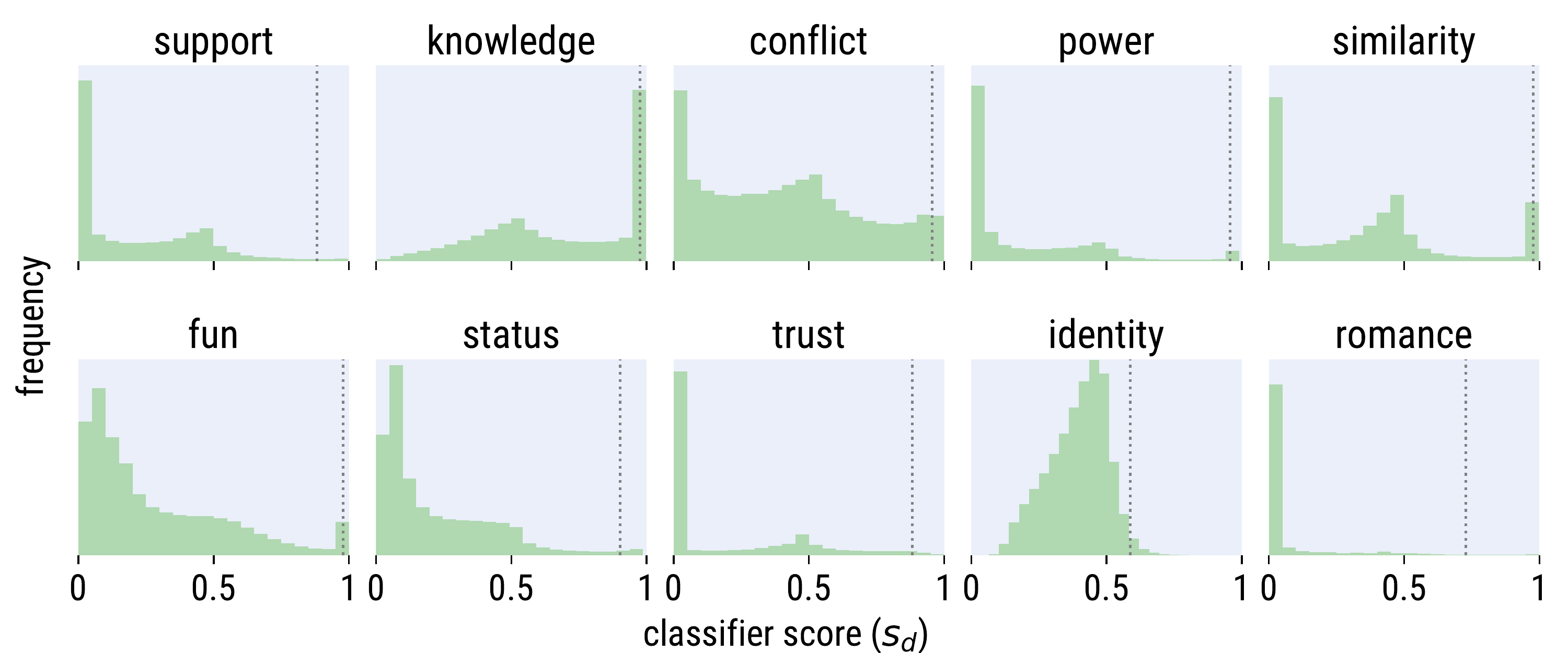}
    \includegraphics[width=0.35\linewidth]{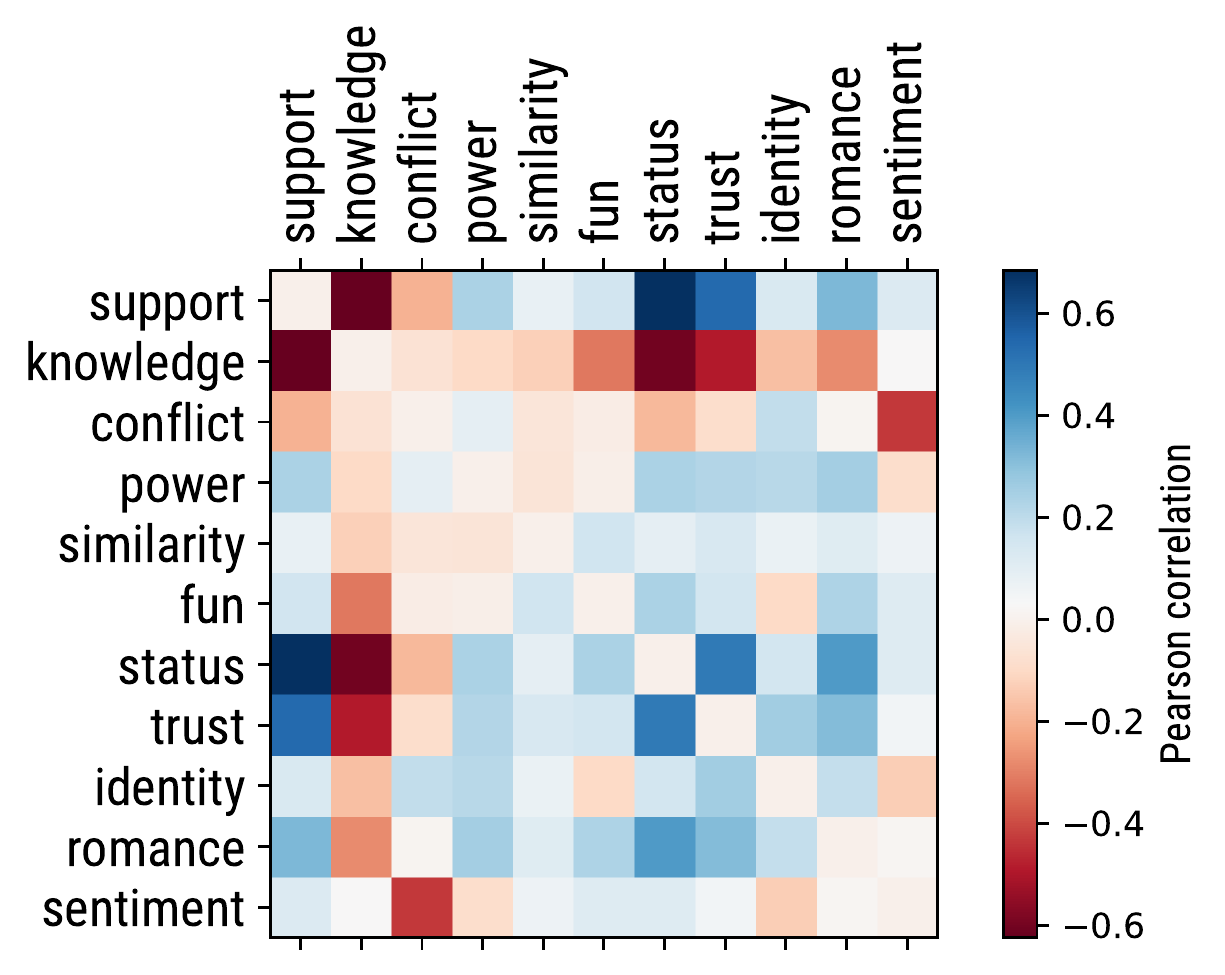}
		\caption{Left: frequency distributions of the classifier scores $s_d$ for all dimensions. The dotted vertical lines mark the values of the $99^{th}$ percentile of each distribution. Right: cross-correlation matrix of the classifier scores of all dimension pairs across all messages, plus a simple measure of text sentiment.}
    \label{fig:cross_corr}
\end{figure}

To conduct our analysis, we binarized the classifier scores $s_d(m)$ using an indicator function that assigns dimension $d$ to $m$ if $s_d(m)$ is above a certain threshold $\theta_d$: 
\begin{equation}
    d(m) = 
    \begin{cases}
      1, & \text{if}\ s_d(m) \geq \theta_d\\
      0, & \text{otherwise}
    \end{cases}
 \end{equation}
We used dimension-specific thresholds because the empirical distribution of the classifier scores $s_d$ varies noticeably across dimensions (see Figure~\ref{fig:cross_corr}, left), which makes the use of a fixed common threshold unpractical. We made a very conservative choice of $\theta_d$ as the value of the $99^{th}$ percentile of the distribution of the classifier score $s_d$, thus favoring high precision over recall. This effectively reduces the number of messages to 1\% of the total and the number of edges to slightly more than 1\% of the total. In Supplementary Information (Figure~SI2, right), we experimented with different percentiles, starting from the $75^{th}$.

As a result of this procedure, a comment could end up being labeled with multiple dimensions. To measure the extent to which pairs of dimensions are related, we computed the Spearman rank cross-correlation matrix of the classifier scores of all dimension pairs across all messages (Figure~\ref{fig:cross_corr}, right). Some pairs of dimensions such as \emph{status}, \emph{trust} and \emph{support} occur more frequently together, but overall the ten dimension model exhibits a fairly high degree of orthogonality. To make sure that the ten dimension classifier is not capturing simply the sentiment of the text, we correlated the dimensions scores with the scores from Vader, a simple yet widely-used sentiment analyzer~\cite{hutto2014vader}. The correlations were all very low except for a negative correlation with the conflict dimension.

\subsection*{Communication graphs}\label{sec:dataset:reddit_dimensions}

\begin{figure}[t!]
    \centering
    \includegraphics[width=0.60\linewidth]{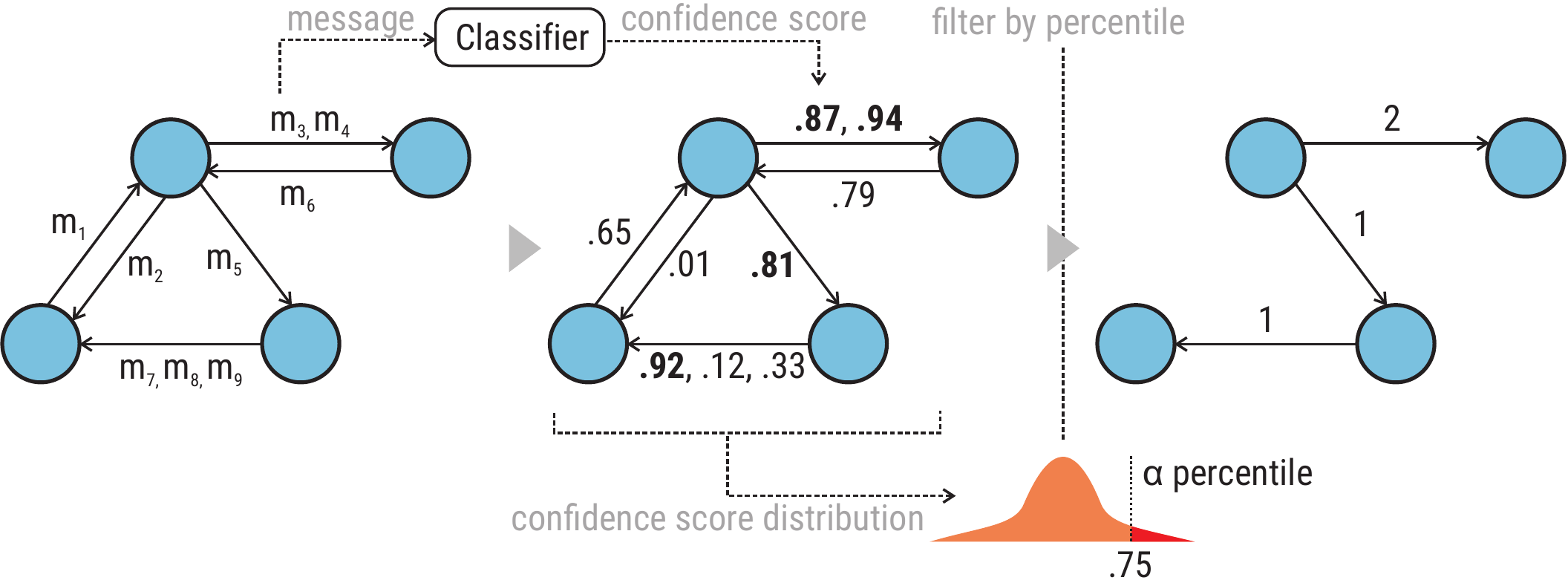}\\
		\caption{Example of how a dimension-specific conversation multigraph $\mathcal{G}_d$ is built. First, the text classifier for dimension $d$ is applied to all messages and outputs scores that are proportional to the likelihood of a message containing dimension $d$. Then, for each dimension individually, a score threshold is determined based on a selected percentile $\alpha$ in the overall score distribution. In the illustrated example, the value corresponding to the $\alpha$ percentile is $0.75$. Last, only the edges with the messages that pass that threshold are kept; the messages are counted to compute the edge weight.}
    \label{fig:graph_filtering}
\end{figure}

In Reddit, conversations develop over discussion threads. If user $i$ commented over either a submission or a comment of another user $j$, we considered that $i$ sent a \emph{message} to $j$. We created a directed communication graph $\mathcal{G}(U,E)$ to model such exchange of messages. The set of nodes $U$ contains all the geo-referenced users in our sample. We connected two users $i$ and $j$ with a directed edge $(i, j, w(i, j)) \in E$ if user $i$ sent at least one message to user $j$. The edge weight $w(i, j)$ represents the ties strength and it is equal to the total number of messages sent. Enforcing a minimum threshold on edge weights for them to be included in the communication graph improved the results, likely because it filters out ``occasional'' interactions that do not provide a strong signal about the type of social relationships. We used the optimal threshold of $w(i, j) \geq 4$; in Supplementary Information (Figure~SI2, left) we present results with different thresholds.

By labeling each message according to the ten social dimensions, we could extract dimension-specific conversation graphs $\mathcal{G}_d$, namely a subgraph of $\mathcal{G}$ created using only the messages that contain dimension $d$. We built such subgraph using the procedure illustrated in Figure~\ref{fig:graph_filtering}. Given a message $m$, we computed its classifier score $s_d(m)$, which is proportional to the likelihood of $m$ containing expressions of dimension $d$. We kept only the messages whose likelihood is higher than a dimension-specific threshold: $s_d(m) \geq \theta_d$. In practice, we assigned to $\theta_d$ the value of the $99^{th}$ percentile of the empirical distribution of $s_d(m)$ values, which effectively retains only $1\%$ of the messages. Given such a heavy filtering, we did not enforce a threshold on edge weights. Based on this reduced sets of messages, we constructed a new dimension-specific graph $\mathcal{G}(U_d,E_d)$ that was effectively a subgraph of the original communication graph where an edge $(i, j, w_d(i, j)) \in E_d$ encoded the fact that user $i$ sent $w_d(i, j)$ messages conveying dimension $d$ to user $j$. When messages were labeled with multiple dimensions, they contributed equally to multiple dimension-specific subgraphs.

\subsection*{Computing diversity of interactions}\label{sec:dataset:diversity}

Eagle et al.~\cite{eagle2010network} define two measures of diversity: social $D_{social}$ and spatial $D_{spatial}$. In practice, the two metrics are highly correlated, hence in the main Results we report findings for $D_{spatial}$. In Supplementary Information, we discuss findings for both diversity measures.

Given a user $i$, we first calculated the proportion of the total number of messages that $i$ sent to $j$, namely:
\begin{equation}
p_{ij} = \frac{w(i,j)} {\sum_{j=1}^{k} w(i,j)},
\label{eqn:pij}
\end{equation}
where $k$ is the total number of $i$'s social contacts on the communication graph $\mathcal{G}$. In telephone network, the strength of a tie was measured as the total call duration, whereas we measured it as the total number of messages. We then calculated the normalized Shannon entropy of those proportions:
\begin{equation}
D_{social}(i) = \frac{-\sum_{j=1}^{k}  p_{ij} \cdot  \log(p_{ij})} {\log(k) }.
\label{eqn:dsocial}
\end{equation}
The dimension-specific social diversity was computed with an analogous formula, but taking into account only the edges in the dimension-specific graph $\mathcal{G}_d$:
\begin{equation}
p^d_{ij} = \frac{w_d(i,j)} {\sum_{j=1}^{k_d} w_d(i,j)},
\label{eqn:pijd}
\end{equation}
\begin{equation}
D^{d}_{social}(i) = \frac{-\sum_{j=1}^{k_d}  p^d_{ij} \cdot  \log(p^d_{ij})} {\log(k_d) },
\label{eqn:dsociald}
\end{equation}
where $k_d$ is the total number of $i$'s social contacts on the dimension-specific graph $\mathcal{G}_d$. To compute the spatial diversity $D_{spatial}$, we first calculated the proportion of total volume of messages exchanged by user $i$ with any other users living in area $a$:
\begin{equation}
p_{ia} = \frac{ \sum_{j \in U_a} w(i,j)  } {\sum_{j=1}^{k} w(i,j)},
\label{eqn:pia}
\end{equation}
where $A$ is the total number of areas and $U_a \subset U$ is the subset of users living in area $a$. We then computed the spatial diversity as the normalized entropy of the $p_{ia}$ proportions:
\begin{equation}
D_{spatial}(i) = \frac{-\sum_{a=1}^{A}  p_{ia} \cdot  \log(p_{ia})} {\log(A) }.
\label{eqn:dspatial}
\end{equation}
The same formulation is applied to the dimension-specific graphs:
\begin{equation}
p^{d}_{ia} = \frac{ \sum_{j \in U_a} w_{d}(i,j)  } {\sum_{j=1}^{k_d} w_{d}(i,j)},
\label{eqn:pia_d}
\end{equation}
\begin{equation}
D^{d}_{spatial}(i) = \frac{-\sum_{a=1}^{A}  p^{d}_{ia} \cdot  \log(p^{d}_{ia})} {\log(A) }.
\label{eqn:dspatial_d}
\end{equation}
Last, we computed the diversity values at area level by averaging the diversity scores of users living in the same area:
\begin{equation}
D_{social}(a) = \frac{\sum_{i \in U_a} D_{social}(i)}{|U_a|}; D^{d}_{social}(a) = \frac{\sum_{i \in U_a} D^d_{social}(i)}{|U_a|}
\label{eqn:dsocial_area}
\end{equation}
\begin{equation}
D_{spatial}(a) = \frac{\sum_{i \in U_a} D_{spatial}(i)}{|U_a|}; D^{d}_{spatial}(a) = \frac{\sum_{i \in U_a} D^d_{spatial}(i)}{|U_a|}
\label{eqn:dspatial_area}
\end{equation}

\subsection*{Linear regression}

Linear regression is an approach for modeling a linear relationship between a dependent variable (GDP, in our experiments) and a set of independent variables (diversity measures), and it does so by associating a so-called $\beta$-coefficient with each independent variable such as the sum of all independent variables multiplied by their respective $\beta$-coefficients approximates the value of the dependent variable with minimal error. Specifically, we used an Ordinary Least Squares (OLS) regression model to estimate the coefficients such that the sum of the squared residuals between the estimation and the actual value is minimized. The diversity metrics given in input to the regression were approximately normally distributed and bounded in the interval [0,1] (see Figure~SI5)

\subsection*{Modeling geographical span}\label{sec:distance:vs_dist}

To study the dependency between geographical space and social dimensions, we estimated the conditional probability $p(d|l)$ of a dimension $d$ occurring in conversations characterized by a given \emph{geographic span} (or length) $l$. Specifically, we considered the set $E@l$ of all edges in the conversation graph $\mathcal{G}$ that connect users at geographic distance $l$, and the subset of those edges $E_d@l$ that belong to the dimension-specific graph $\mathcal{G}_d$. We then computed the conditional probability as the number of dimension-specific edges over the total number of edges at distance $l$, namely: $p(d|l) = \frac{|E_d@l|}{|E@l|}$.

Because activity and connectivity are not uniformly distributed across states, the probability $p(d|l)$ alone could yield a biased view of the interplay between interactions and space. To understand why, consider a scenario in which most of the users are concentrated in one single state. In such a scenario, all users would be constrained to interact mostly with people from that state, and the resulting spatial patterns will be just reflecting the underlying activity and spatial distributions rather than being indicative of explicit user choices. To account for this, we discounted $p(d|l)$ by a probability $p_{null}(d|l)$ computed on randomized data. In particular, we generated a random \emph{null} model by randomly reshuffling the locations across users. By doing so, we preserved both the connectivity properties of the conversation network and the population distribution across states, yet destroying the original relationship between social links and spatial locations. Finally, we computed a normalized score $\Delta p(d|l) = \frac{p(d|l)}{p_{null}(d|l)} -1$, which measures the \% change of the probability of interaction compared to what it is expected by chance. To obtain the conditional probability associated to individual messages rather than social links, we also computed an alternative version of $\Delta p(d|l)$ that considers each message as an individual edge in the graph, thus effectively weighting more pairs of individuals who communicated often.

Since we could geo-reference users at state-level only, we approximated the span of a social link between two users to the length of the straight line connecting the geographic centroids of their states. Given the relatively limited spatial resolution of such a definition, we were bound to a coarse partitioning of distances. Effectively, we divided the set of edges in quintiles based on their geographic span distribution, thus obtaining five equally-sized distance bins, the first of which contains almost exclusively interactions among people in the same state ($l=0$).

\section*{Data and Code Availability}

We made all the data used in this study publicly available. The data consists of: 1) individual messages scored with the ten dimension classifier and the identifiers of the sender and receiver; 2) estimated location of the users in the communication graph; 3) aggregated data at state-level reporting the diversity metrics. The DOI of the publicly accessible data is \url{10.6084/m9.figshare.19918231}. The pre-trained social dimensions classifier is available at \url{http://www.github.com/lajello/tendimensions}.

\section*{Acknowledgements}

LMA acknowledges the support from the Carlsberg Foundation through the COCOONS project (CF21-0432). The funder had no role in study design, data collection and analysis, decision to publish, or preparation of the manuscript.

\section*{Author contributions}

LMA conceived the experiments, conducted the analysis, and wrote the manuscript. SJ collected the data and revised the manuscript. DQ conceived the experiments and wrote the manuscript. 

\section*{Competing interests}

The authors declare no competing interests.

\newpage

\section*{Supplementary Information to ``Multidimensional Tie Strength and Economic Development''}

\setcounter{figure}{0}
\setcounter{table}{0}

\renewcommand{\thefigure}{SI\arabic{figure}}
\renewcommand{\thetable}{SI\arabic{table}}

\subsection*{Network distributions}

\begin{figure}[ht!]
    \centering
		\includegraphics[width=0.64\linewidth]{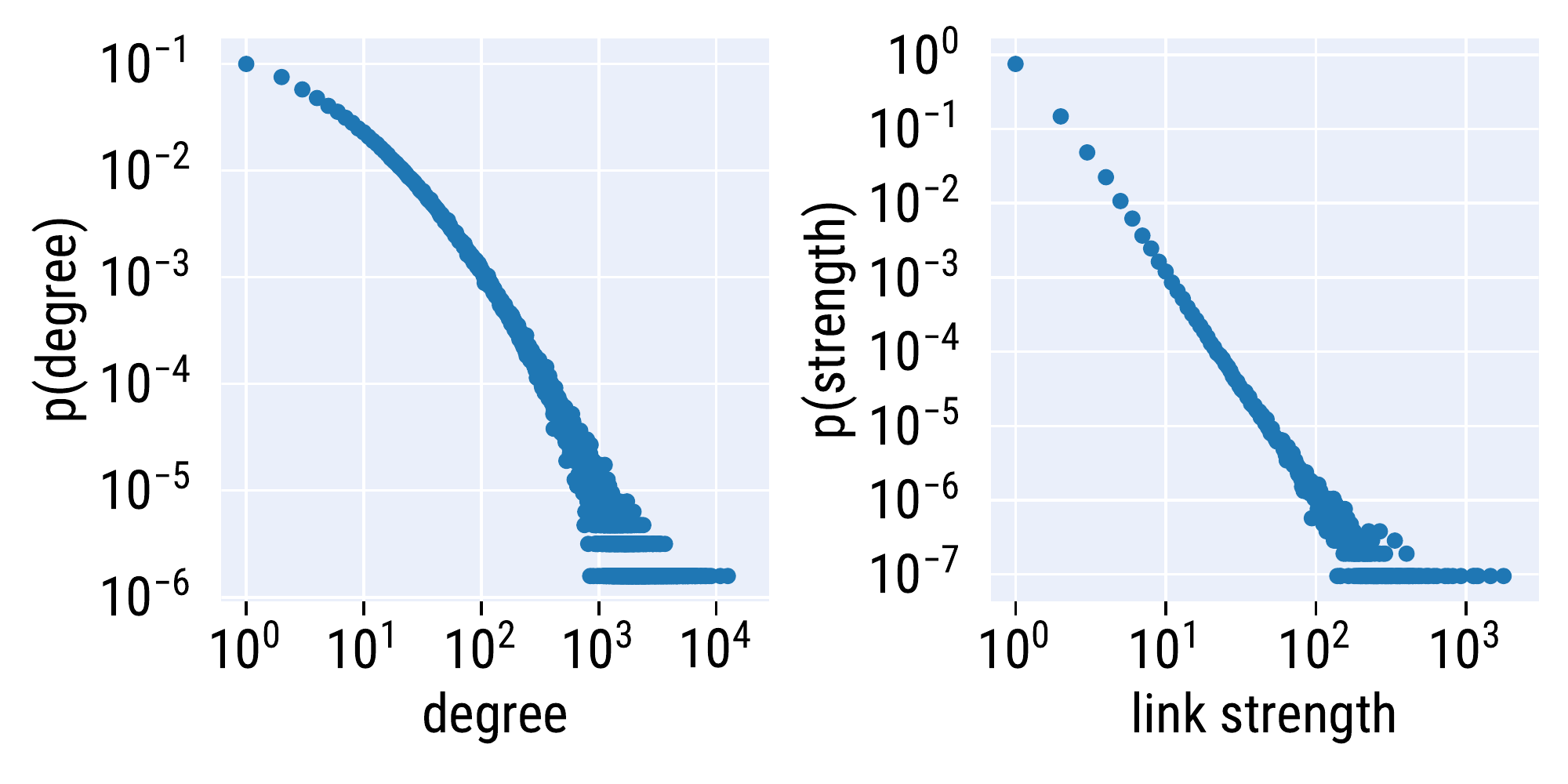}
		\caption{Distributions of node degree (left) and link strength calculated as the number of messages flowing between connected nodes (right) in the communication graph $\mathcal{G}$.}
    \label{fig:distributions}
\end{figure}
The distributions of the degree and strength of the Reddit communication network are shown in Figure~\ref{fig:distributions}.

\subsection*{Regressions without population density}

{\def\arraystretch{1.5}
\begin{table}[!h]
\footnotesize
\setlength{\tabcolsep}{5pt}
\begin{center}
\textbf{Predicting GDP per capita from:}\\
\quad
\begin{tabular}{lccc}
\specialrule{.1em}{.05em}{.05em} 
\multicolumn{4}{c}{\textbf{Diversity on full communication graph}}\\
\textbf{Feature}	& \textbf{$\beta$} &	\textbf{SE} & \textbf{$p$} \\
\hline
$\alpha$ (intercept)	           & 0.182  & 0.093  & 0.058 \\
$D_{spatial}$                    & 0.436 & 0.154  & 0.007 \\
&&&\\
\hline
\multicolumn{2}{l}{Durbin-Watson stat. = 2.280} 	& \multicolumn{2}{l}{\textbf{$R_{adj}^2$ = 0.14}}   \\
\specialrule{.1em}{.05em}{.05em} 
\end{tabular}
\quad
\begin{tabular}{lccc}
\specialrule{.1em}{.05em}{.05em} 
\multicolumn{4}{c}{\textbf{Spatial diversity on dimension-specific graphs}}\\
\textbf{Feature}	& \textbf{$\beta$} &	\textbf{SE} & \textbf{$p$} \\
\hline
$\alpha$ (intercept)	           & 0.244  & 0.070  & 0.001 \\
$D_{spatial}^{knowledge}$         & 1.168  & 0.188  & 0.000 \\
$D_{spatial}^{support}$          & -0.593 & 0.180  & 0.002 \\
\hline
\multicolumn{2}{l}{Durbin-Watson stat. = 1.983} 	& \multicolumn{2}{l}{\textbf{$R_{adj}^2$ = 0.48}}   \\
\specialrule{.1em}{.05em}{.05em} 
\end{tabular}
\end{center}
\caption{Linear regressions to predict GDP per capita of US states from: (left) spatial diversity computed on the full communication graph; (right) spatial diversities computed on dimension-specific communication graphs. Adjusted $R^2$ and Durbin-Watson statistic for autocorrelation (values close to 2 indicate no autocorrelation) are reported. The contribution of individual features to the models is described by their $beta$-coefficients, standard errors (SE) and $p$-values.}
\label{tab:regression_nodensity}
\end{table}
}

Table~\ref{tab:regression_nodensity} shows the results of regression models that include spatial diversity but exclude population density as a control variable. For reference, a model that predicts GDP from population density only achieved an $R_{adj}^2$ of 0.26.

\subsection*{Spatial vs. social diversity}

{\def\arraystretch{1.5}
\begin{table}[ht!]
\footnotesize
\setlength{\tabcolsep}{5pt}
\begin{center}
\textbf{Predicting GDP per capita from:}\\
\begin{tabular}{lccc}
\specialrule{.1em}{.05em}{.05em} 
\multicolumn{4}{c}{\textbf{Social+Spatial diversity on full graph}}\\
\textbf{Feature}	& \textbf{$\beta$} &	\textbf{SE} & \textbf{$p$} \\
\hline
$\alpha$ (intercept)	           & 0.065  & 0.094  & 0.493 \\
Pop. density                     & 0.581  & 0.159  & 0.001 \\
$D_{social}$                     & 0.489  & 0.190  & 0.014 \\
$D_{spatial}$                    & -0.031 & 0.179  & 0.864 \\
\hline
\multicolumn{2}{l}{Durbin-Watson stat. = 2.108} 	& \multicolumn{2}{l}{\textbf{$R_{adj}^2$ = 0.38}}   \\
\specialrule{.1em}{.05em}{.05em} 
\end{tabular}
\begin{tabular}{lccc}
\specialrule{.1em}{.05em}{.05em} 
\multicolumn{4}{c}{\textbf{Social diversity on dimension-specific graphs}}\\
\textbf{Feature}	& \textbf{$\beta$} &	\textbf{SE} & \textbf{$p$} \\
\hline
$\alpha$ (intercept)	           & 0.193  & 0.066  & 0.006 \\
Pop. density                     & 0.440  & 0.121  & 0.001 \\
$D_{social}^{knowledge}$                     & 0.884  & 0.147  & 0.000 \\
$D_{social}^{support}$                     & -0.435 & 0.142  & 0.004 \\
\hline
\multicolumn{2}{l}{Durbin-Watson stat. = 2.131} 	& \multicolumn{2}{l}{\textbf{$R_{adj}^2$ = 0.60}}   \\
\specialrule{.1em}{.05em}{.05em} 
\end{tabular}
\\
\quad
\begin{tabular}{lccc}
\multicolumn{4}{c}{\textbf{ }}\\
\specialrule{.1em}{.05em}{.05em} 
\multicolumn{4}{c}{\textbf{Social+spatial on dimension-specific graphs}}\\
\textbf{Feature}	& \textbf{$\beta$} &	\textbf{SE} & \textbf{$p$} \\
\hline
$\alpha$ (intercept)	           & 0.190  & 0.059  & 0.003 \\
Pop. density                     & 0.492  & 0.113  & 0.000 \\
$D_{social}^{knowledge}$         & 1.052  & 0.159  & 0.000 \\
$D_{spatial}^{support}$          & -0.585 & 0.152  & 0.000 \\
\hline
\multicolumn{2}{l}{Durbin-Watson stat. = 2.110} 	& \multicolumn{2}{l}{\textbf{$R_{adj}^2$ = 0.64}}   \\
\specialrule{.1em}{.05em}{.05em} 
\end{tabular}
\quad
\begin{tabular}{lccc}
\multicolumn{4}{c}{\textbf{ }}\\
\specialrule{.1em}{.05em}{.05em} 
\multicolumn{4}{c}{\textbf{Social+spatial on dimension-specific graphs}}\\
\textbf{Feature}	& \textbf{$\beta$} &	\textbf{SE} & \textbf{$p$} \\
\hline
$\alpha$ (intercept)	           & 0.183  & 0.069  & 0.012 \\
Pop. density                     & 0.446  & 0.127  & 0.001 \\
$D_{spatial}^{knowledge}$         & 0.8081 & 0.148  & 0.000 \\
$D_{social}^{support}$          & -0.339 & 0.141  & 0.021 \\
\hline
\multicolumn{2}{l}{Durbin-Watson stat. = 2.027} 	& \multicolumn{2}{l}{\textbf{$R_{adj}^2$ = 0.56}}   \\
\specialrule{.1em}{.05em}{.05em} 
\end{tabular}

\end{center}
\caption{Linear regressions to predict GDP per capita of US states from social diversity in the full graph (top, left), social diversity computed on dimension-specific communication graphs (top, right), and combinations of social and spatial diversity on the on dimension-specific communication graphs (bottom), left and right. Population density is added as a control variable. Adjusted $R^2$ and Durbin-Watson statistic for autocorrelation (values close to 2 indicate no autocorrelation) are reported. The contribution of individual features to the models is described by their $beta$-coefficients, standard errors (SE) and $p$-values.}
\label{tab:regression_uniform_type}
\end{table}
}

Table~\ref{tab:regression_uniform_type} reports the results of linear regressions that include social diversity, alone or combined with spatial diversity. The combination of $D^{knowledge}_{social}$ and $D^{support}_{spatial}$ yields the best fit ($R^2_{adj} = 0.64$), only slightly above the model that considers only spatial diversity ($R^2_{adj} = 0.62$ Table~1).

\subsection*{Sensitivity to minimum edge weight and classifier threshold}

\begin{figure}[ht!]
    \centering
		\includegraphics[width=0.60\linewidth]{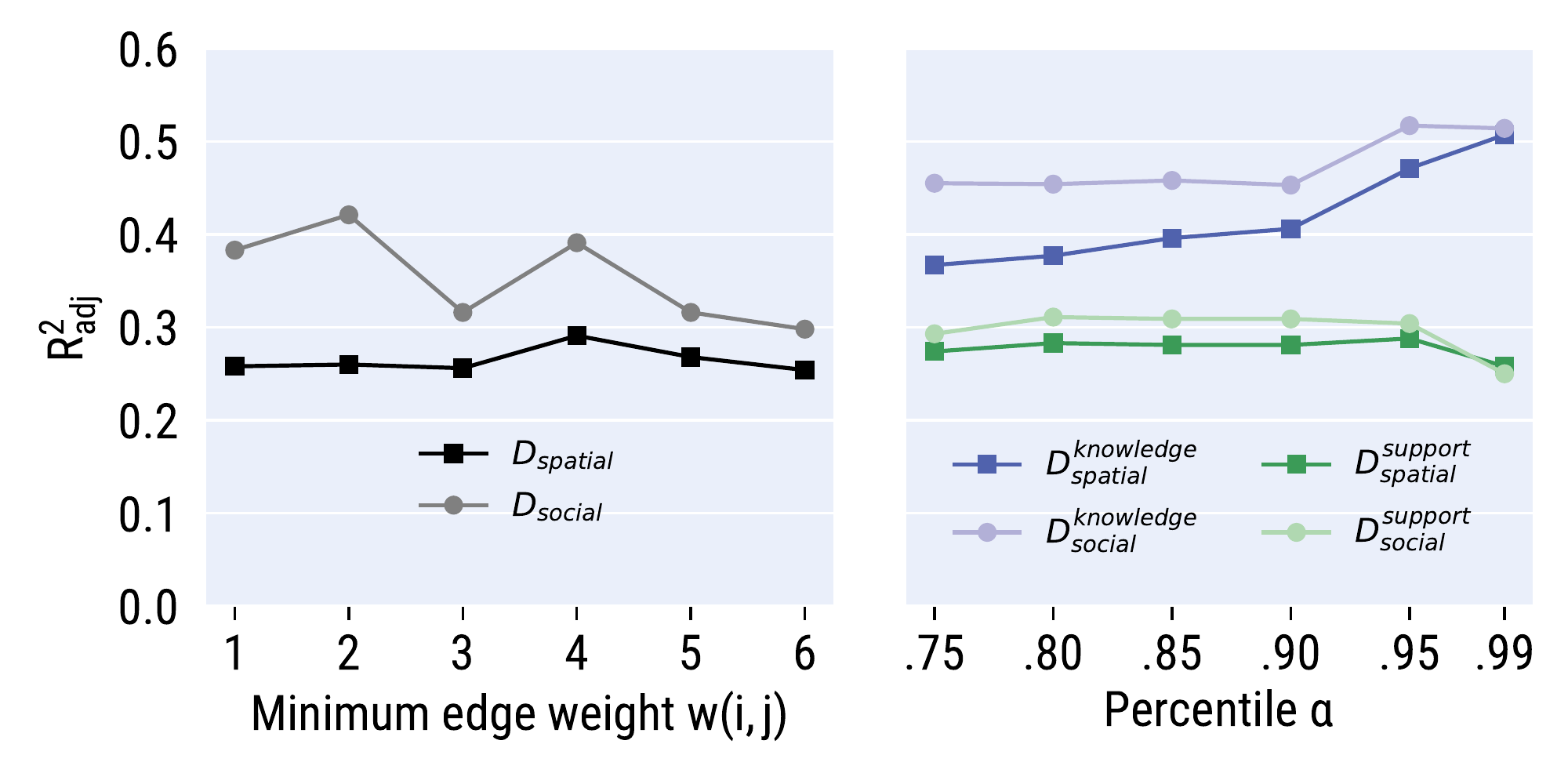}
		\caption{Left: adjusted $R^2$ of linear regressions that predict GDP from diversity on the full graph ($D_{spatial}$, $D_{social}$) as the minimum weight $w$ of edges included in the graph varies. Right: adjusted $R^2$ of univariate linear regressions to predict GDP from social and spatial diversity on the \emph{knowledge} graph and \emph{support} graphs as the percentile $\alpha$ used to binarize the classifier scores varies. We included population density as control variable in all regressions.}
    \label{fig:sensitivity_alpha}
\end{figure}

We included in the conversation graph $\mathcal{G}$ only edges with minimum weight of 4. This was the optimal threshold we found in the range $[1,6]$ (Figure~\ref{fig:sensitivity_alpha}, left). Similarly, the dimension-specific graphs $\mathcal{G}_d$ are obtained after setting a threshold $\theta_d$ equal to the value of the $\alpha$ percentile on the distribution of the classifier scores $s_d$. We explored different values of $\alpha$ and found that the $99^{th}$ percentile works best (Figure~\ref{fig:sensitivity_alpha}, right). However, dimension-specific graphs created using any percentiles above the $75^{th}$ yielded better prediction than any version of the full conversation graph.

\subsection*{Sensitivity to set of states}

{\def\arraystretch{1.5}
\begin{table}[ht!]
\footnotesize
\setlength{\tabcolsep}{5pt}
\begin{center}

\begin{tabular}{lccc}
\multicolumn{4}{c}{\textbf{ }}\\
\specialrule{.1em}{.05em}{.05em} 
\multicolumn{4}{c}{\textbf{Population density}}\\
\textbf{Feature}	& \textbf{$\beta$} &	\textbf{SE} & \textbf{$p$} \\
\hline
$\alpha$ (intercept)	           & 0.393  & 0.042  & 0.000 \\
Pop. density                     & 0.419  & 0.154  & 0.009 \\
&  &  &  \\
&  &  &  \\
\hline
\multicolumn{2}{l}{Durbin-Watson stat. = 2.150} 	& \multicolumn{2}{l}{\textbf{$R_{adj}^2$ = 0.12}}   \\
\specialrule{.1em}{.05em}{.05em} 
\end{tabular}
\quad
\begin{tabular}{lccc}
\multicolumn{4}{c}{\textbf{Predicting GDP per capita from:}}\\
\specialrule{.1em}{.05em}{.05em} 
\multicolumn{4}{c}{\textbf{Diversity on full communication graph}}\\
\textbf{Feature}	& \textbf{$\beta$} &	\textbf{SE} & \textbf{$p$} \\
\hline
$\alpha$ (intercept)	           & 0.264  & 0.093  & 0.007 \\
Pop. density                     & 0.372  & 0.164  & 0.028 \\
$D_{social}$                     & 0.271  & 0.181  & 0.142 \\
$D_{spatial}$                    & -0.014 & 0.174  & 0.934 \\
\hline
\multicolumn{2}{l}{Durbin-Watson stat. = 2.175} 	& \multicolumn{2}{l}{\textbf{$R_{adj}^2$ = 0.13}}   \\
\specialrule{.1em}{.05em}{.05em} 
\end{tabular}
\quad
\begin{tabular}{lccc}
\multicolumn{4}{c}{\textbf{ }}\\
\specialrule{.1em}{.05em}{.05em} 
\multicolumn{4}{c}{\textbf{Diversity on dimension-specific graphs}}\\
\textbf{Feature}	& \textbf{$\beta$} &	\textbf{SE} & \textbf{$p$} \\
\hline
$\alpha$ (intercept)	           & 0.499  & 0.101  & 0.000 \\
Pop. density                     & 0.303  & 0.142  & 0.039 \\
$D_{social}^{knowledge}$         & 0.661  & 0.171  & 0.000 \\
$D_{spatial}^{support}$          & -0.555 & 0.202  & 0.009 \\

\hline
\multicolumn{2}{l}{Durbin-Watson stat. = 2.339} 	& \multicolumn{2}{l}{\textbf{$R_{adj}^2$ = 0.30}}   \\
\specialrule{.1em}{.05em}{.05em} 
\end{tabular}

\end{center}
\caption{
Linear regressions to predict GDP per capita of 50 US states from: (left) population density only; (center) spatial and social diversity computed on the full communication graph; (right) spatial and social diversity computed on dimension-specific communication graphs. The variables $D^{knowledge}_{social}$ and $D^{support}_{spatial}$ were picked automatically by a feature-selection algorithm out of all the dimension-specific diversity measures. Population density is added as a control variable in the latter two models. Adjusted $R^2$ and Durbin-Watson statistic for autocorrelation (values close to 2 indicate no autocorrelation) are reported. The contribution of individual features to the models is described by their $beta$-coefficients, standard errors (SE) and $p$-values.}
\label{tab:regression_all_states}
\end{table}
}

We restricted our study to a subset of 44 states, after removing states whose Reddit penetration was too low or was not proportional to the population of residents. The inclusion of outlier states is detrimental to the performance of all models, yet the quality of fit of the \emph{knowledge}-specific model is far superior to the one obtained using the full graph---two to three times as good. Multivariate regressions with all 50 states are presented in Table~\ref{tab:regression_all_states}; their predictive power is roughly halved compared to that of the models fit on 44 states only. Yet, the dimension-specific model was still 130\% more accurate than the one using diversity computed on the full conversation graph.

\subsection*{Baseline with random selection of links}

We calculated univariate regressions based on a randomized model. Specifically, we predicted GDP from diversity measures calculated on a communication graph created from a random selection of 1\% of the messages. This is equivalent of creating a null model where the \emph{knowledge} or \emph{support} labels of messages are reshuffled at random, such that the association between social links and social dimensions is disrupted. We repeated the experiment for 50 random runs. For $D^{random}_{spatial}$, we obtained $\bar{R}_{adj}^2 = 0.084$,~$(stdev = 0.090)$; for $D^{random}_{spatial}$, we obtained $\bar{R}_{adj}^2 = 0.096$,~$(stdev = 0.101)$. The $R^2$ of these random models are much lower than those obtained considering 1\% of \emph{knowledge} or \emph{support} messages (${R}_{adj}^2 = 0.35$ and ${R}_{adj}^2 = 0.26$, respectively).

\subsection*{Sensitivity to temporal window}

\begin{figure}[ht!]
    \centering
		\includegraphics[width=0.80\linewidth]{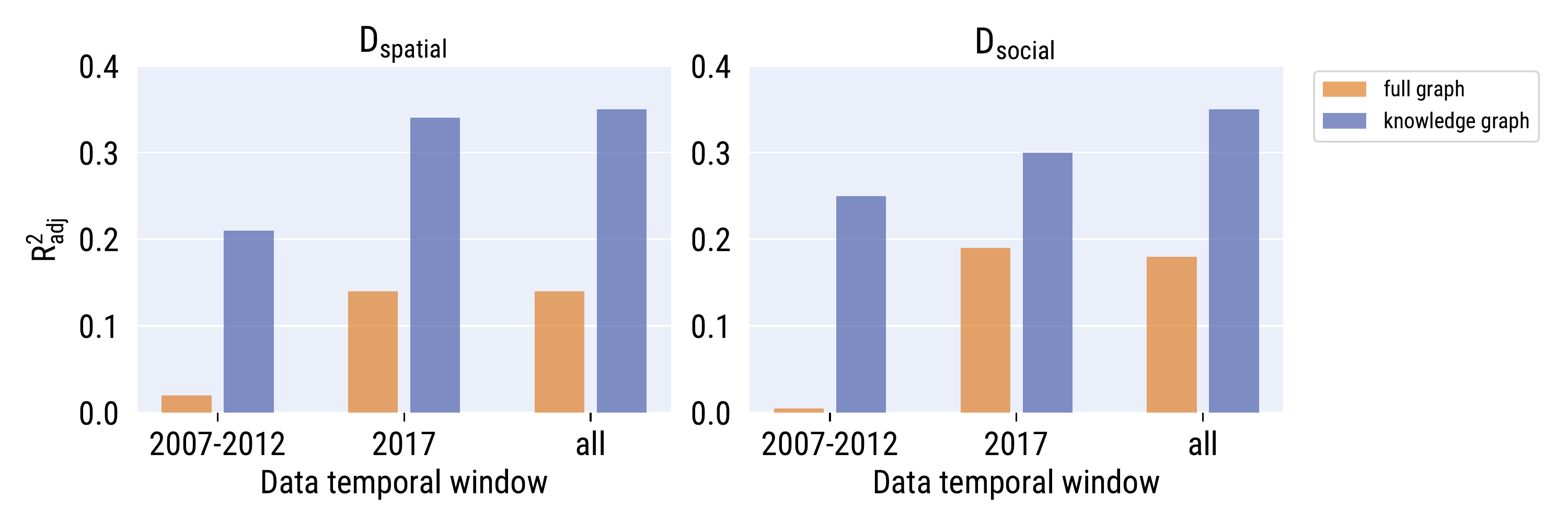}
		\caption{Sensitivity of data temporal window. Adjusted $R^2$ of univariate linear regressions to predict GDP from social and spatial diversity from three temporal windows: 2007 to the end of 2012, 2017, and the full dataset. Results for the full conversation graph and for the knowlegde graph are compared.}
    \label{fig:sensitivity_temporal_window}
\end{figure}

The Reddit data we collected comes from two distinct time periods: from 2007 to the end of 2012 and during the whole year of 2017. In our study, we built conversation graphs with the data from these two periods jointly (all). Later, we explored how conversation graphs made from each period individually are predictive of GDP in year 2017. Figure~\ref{fig:sensitivity_temporal_window} shows the adjusted $R^2$ of univariate linear regressions to predict GDP from social and spatial diversity, using either the full conversation graph or the \emph{knowledge} graph. The most recent data from year 2017 best approximated the results obtained using all the data. The performance decay in the earliest temporal window of years 2007-2012 was conspicuous for models based on the full graph: their $R^2$ dropped close to zero. The performance decay should not be attributed to data sparsity because, in our dataset, posts published in the period 2007-2012 are more abundant than those published in 2017 (9.8M vs. 6.4M).

This result suggests that \emph{knowledge}-specific interactions not only better predict economic development; they also provide a predictive signal that is more resilient to temporal shifts of the data relative to the time in which the outcome variable was measured.

\subsection*{Sensitivity to threshold for user geo-referencing}

\begin{figure}[ht!]
    \centering
		\includegraphics[width=0.80\linewidth]{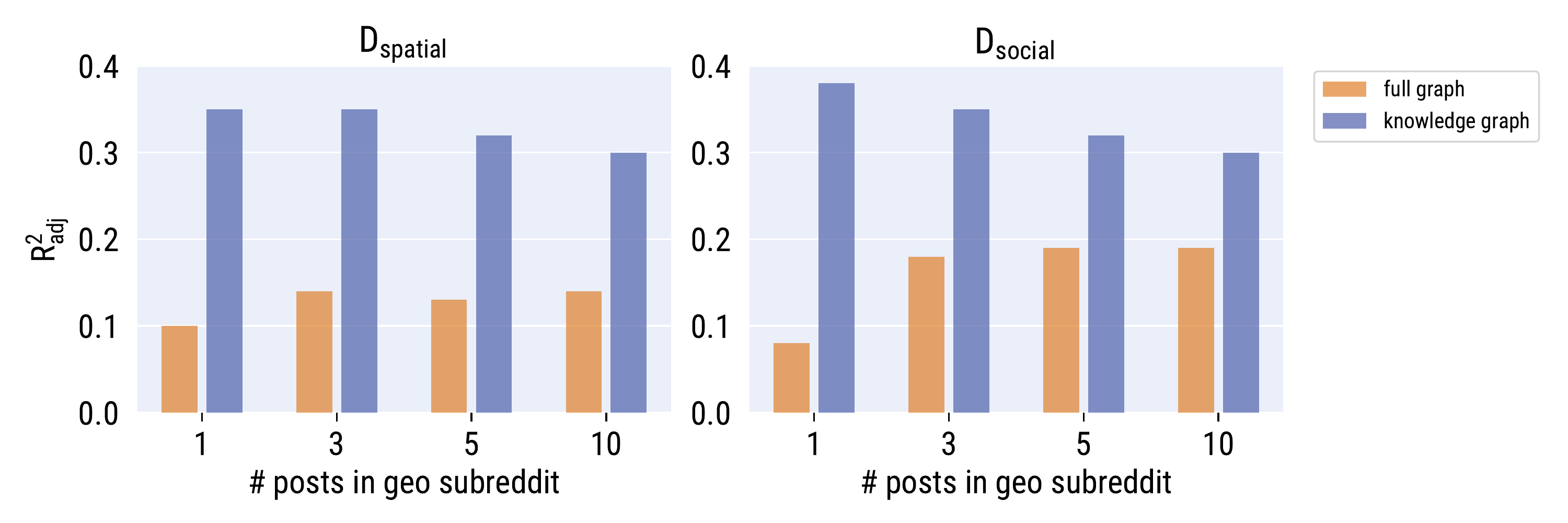}
		\caption{Adjusted $R^2$ of regression models whose variables are obtained from conversation graphs including different sets of geo-referenced users. Specifically, we included only users who posted a minimum number of comments or submissions in geo-salient subreddits. Results for the full conversation graph and for the knowlegde graph are compared.}
    \label{fig:sensitivity_mincount}
\end{figure}

To geo-reference Reddit users, we analyzed their activity in geo-salient subreddits. Specifically, we assigned a user to a US state if they posted at least 3 submissions or comments in subreddits related to that state. We explored how the prediction results change by varying this threshold. Figure~\ref{fig:sensitivity_mincount} shows the adjusted $R^2$ of univariate linear regressions to predict GDP from social and spatial diversity, using either the full conversation graph or the \emph{knowledge} graph. Results are broken down by different thresholds of minimum number of geo-salient posts required for assigning a user to a geographical location. Raising the threshold is beneficial to the model based on the full communication graph, as it helps filtering out incorrect user-state associations. The $R^2$ on the \emph{knowledge} network slightly declines as the threshold increases; this is likely due to the fact that further filtering on a network that contains only 1\% of links aggravates data sparsity.

\subsection*{Distributions of diversity scores }

\begin{figure}[ht!]
    \centering
		\includegraphics[width=0.80\linewidth]{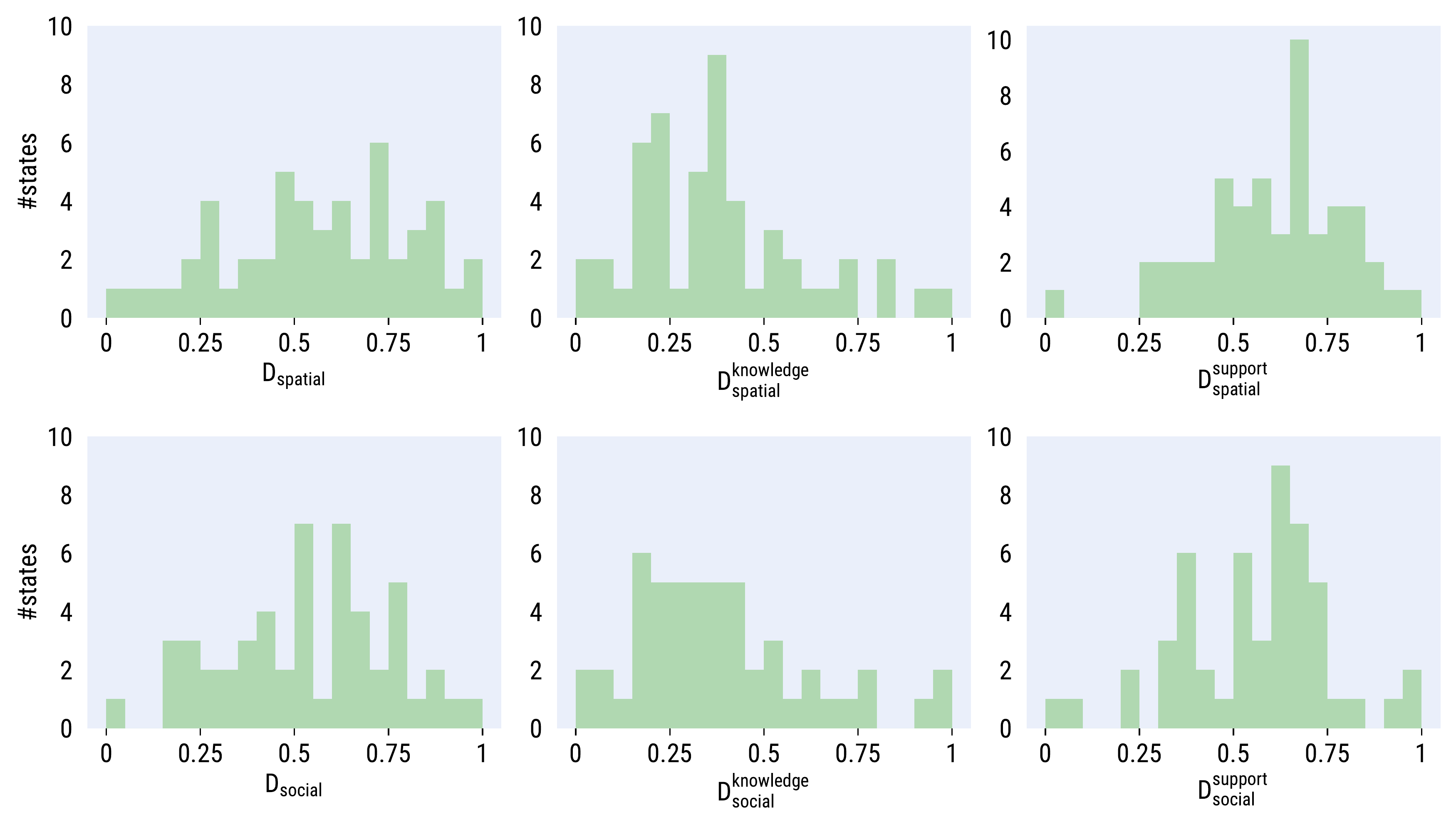}
		\caption{Distribution of diversity scores calculated on the full graph ($D_{spatial}$, $D_{social}$) and on the \emph{knowledge} and \emph{support} graphs ($D^{knowledge}_{spatial}$, $D^{knowledge}_{social}$, $D^{support}_{spatial}$, $D^{support}_{social}$). Values are min-max normalized.}
    \label{fig:distr_diversity}
\end{figure}

In Figure~\ref{fig:distr_diversity}, we report the distribution of diversity scores across US states calculated on the full graph and on the \emph{knowledge} graph.

\subsection*{Multivariate models with alternative sets of variables}

{\def\arraystretch{1.5}
\begin{table}[ht!]
\footnotesize
\setlength{\tabcolsep}{5pt}
\begin{center}

\begin{tabular}{lccc}
\multicolumn{4}{c}{\textbf{Predicting GDP per capita}}\\
\specialrule{.1em}{.05em}{.05em} 
\textbf{Feature}	& \textbf{$\beta$} &	\textbf{SE} & \textbf{$p$} \\
\hline
$\alpha$ (intercept)	          & 0.219  & 0.130 & 0.102 \\
Pop. density                    & 0.471  & 0.164 & 0.007 \\
$D_{social}^{knowledge}$        & 0.748  & 0.510 & 0.153 \\
$D_{social}^{support}$          & -0.445 & 0.279 & 0.120 \\
$D_{social}^{conflict}$         & -0.117 & 0.300 & 0.700 \\
$D_{social}^{status}$           & 0.047  & 0.266 & 0.860 \\
$D_{social}^{power}$            & 0.169  & 0.366 & 0.648 \\
$D_{social}^{trust}$            & -0.074 & 0.264 & 0.780 \\
$D_{social}^{similarity}$       & -0.049 & 0.365 & 0.895 \\
$D_{social}^{identity}$         & 0.214  & 0.329 & 0.520 \\
$D_{social}^{fun}$              & 0.117  & 0.300 & 0.698 \\
$D_{social}^{romance}$          & -0.233 & 0.278 & 0.409 \\
\hline
\multicolumn{2}{l}{Durbin-Watson stat. = 1.998} 	& \multicolumn{2}{l}{\textbf{$R_{adj}^2$ = 0.53}}   \\
\specialrule{.1em}{.05em}{.05em} 
\end{tabular}
\quad
\begin{tabular}{lccc}
\multicolumn{4}{c}{\textbf{Predicting GDP per capita}}\\
\specialrule{.1em}{.05em}{.05em} 
\textbf{Feature}	& \textbf{$\beta$} &	\textbf{SE} & \textbf{$p$} \\
\hline
$\alpha$ (intercept)	           & 0.172  & 0.140 & 0.229 \\
Pop. density                     & 0.487  & 0.157 & 0.004 \\
$D_{spatial}^{knowledge}$         & 0.774  & 0.578 & 0.190 \\
$D_{spatial}^{support}$          & -0.671 & 0.316 & 0.042 \\
$D_{spatial}^{conflict}$         & -0.136 & 0.336 & 0.688 \\
$D_{spatial}^{status}$           & 0.183  & 0.359 & 0.615	\\
$D_{spatial}^{power}$            & 0.050  & 0.391 & 0.900 \\
$D_{spatial}^{trust}$            & -0.096 & 0.302 & 0.752 \\
$D_{spatial}^{similarity}$       & 0.071  & 0.445 & 0.874 \\
$D_{spatial}^{identity}$         & 0.294  & 0.318 & 0.363 \\
$D_{spatial}^{fun}$              & 0.117  & 0.346 & 0.737 \\
$D_{spatial}^{romance}$          & -0.076 & 0.337 & 0.822 \\
\hline
\multicolumn{2}{l}{Durbin-Watson stat. = 1.985} 	& \multicolumn{2}{l}{\textbf{$R_{adj}^2$ = 0.55}}   \\
\specialrule{.1em}{.05em}{.05em} 
\end{tabular}

\end{center}
\caption{Linear regressions to predict GDP per capita of US states from the social and spatial diversity ($D^{d}_{social}$, $D^{d}_{spatial}$) computed on dimension-specific communication graphs. Population density is added as a control variable. Adjusted $R^2$ and Durbin-Watson statistic for autocorrelation (values close to 2 indicate no autocorrelation) are reported. The contribution of individual features to the models is described by their $beta$-coefficients, standard errors (SE) and $p$-values.}
\label{tab:regression_all_vars}
\end{table}
}

\emph{Knowledge} and \emph{support} might not be the only two social dimensions associated to economic development. To systematically evaluate how this association varies when considering a wider set of dimensions, we ran two linear regressions that include the social and spatial diversities of all the social dimensions that our NLP tool can capture, plus population density as control (Table~\ref{tab:regression_all_vars}). The adjusted $R^2$ of these models reach 0.55, which is lower than models considering \emph{knowledge} and \emph{support} only. 

These models include too many features, considering the limited number of datapoints (44 states). As a result, the coefficients of all variables are not statistically significant ($p>0.1$). When the set of independent variables is large, it is common practice to use feature-selection approaches to select only those variables that explain most of the variability of the outcome variable. In our experiments we used \emph{stepAIC}. This method is based on the Akaike Information Criterion, or AIC for short (see Sakamoto et al., \emph{``Akaike information criterion statistics''}, 1986), an estimate of the relative amount of information lost by a model to represent the process that generated the empirical data. The AIC score rewards models that achieve a high goodness-of-fit score and penalizes them if they become overly complex. stepAIC measures the AIC score of models obtained by removing different sets of features from the original model and selects the feature combination that yields the lowest AIC. Automatic feature selection kept two variables: social diversity of \emph{knowledge}, and spatial diversity of \emph{support}. This reduced model (summarized in Table~\ref{tab:regression_uniform_type}) yielded an $R_{adj}^2$ of 0.64, which is only slightly better than the model that considers only the spatial diversity of \emph{knowledge} and \emph{support} (Table~1).

\end{document}